\documentclass[11pt,twoside]{article}

\usepackage{asp2006}
\usepackage{graphicx}

\begin{document}
\title{The Brightest Serendipitous X-ray Sources in ChaMPlane}
\author{Kyle Penner,$^{1}$ Maureen van den Berg,$^{2}$ JaeSub Hong,$^{2}$ Silas Laycock,$^{3}$ Ping Zhao,$^{2}$ and Jonathan Grindlay$^{2}$}
\affil{$^{1}$Department of Astronomy, University of Texas, Austin, TX, USA\\
$^{2}$Harvard--Smithsonian Center for Astrophysics, Cambridge, MA, USA\\
$^{3}$Gemini Observatory, Hilo, HI, USA}

\begin{abstract}
The \emph{Chandra} Multiwavelength Plane (ChaMPlane) Survey is a comprehensive effort to
constrain the population of accretion-powered and coronal low-luminosity X-ray sources
(L$_{X} \la 10^{33}$ erg s$^{-1}$) in the Galaxy.  ChaMPlane incorporates X-ray, optical, and
infrared observations of fields in the Galactic Plane imaged with \emph{Chandra} in the past
six years.
We present the results of a population study of the brightest X-ray sources
in ChaMPlane.  We use X-ray spectral fitting, X-ray lightcurve analysis,
and optical photometry of candidate counterparts to determine the properties of 21 sources.
Our sample includes a previously unreported quiescent low-mass X-ray binary or cataclysmic
variable ($R = 20.9$) and ten stellar sources ($12.5 \la R \la 15$), including one flare
star ($R = 17.3$).
We find that quantile analysis, a
new technique developed for constraining the X-ray spectral properties of low-count sources,
is largely consistent with spectral fitting.
\end{abstract}

\section{Introduction}
Accretion powered binaries with compact objects include white dwarfs in cataclysmic variables (CVs)
and neutron stars/black holes in quiescent low-mass X-ray binaries (qLMXBs).  Space
densities of CVs and qLMXBs are poorly constrained physical quantities due to
small number statistics and systematic selection effects \citep{kp_1995cvs..book.....W}.
The \emph{Chandra} Multiwavelength Plane (ChaMPlane) Survey aims to
constrain the Galactic population of low-luminosity (L$_{X}
\la 10^{33}$ erg s$^{-1}$) accretion-powered X-ray binaries using serendipitous sources detected in
six years of deep ($\ga$ 20 ks) \emph{Chandra} observations.  The current database
includes $\sim$15000 X-ray sources from 122 fields.
ChaMPlane will use \emph{Chandra's} sensitivity to detect CVs and qLMXBs
beyond $\sim 1.2$ kpc, the approximate limit of previous surveys; this future sample
will allow us to study the
spatial and luminosity distributions of CVs and qLMXBs on Galactic scales.

ChaMPlane utilizes deep optical imaging (to $R = 24$) to identify
candidate source counterparts for follow-up studies.
Each of the 122 \emph{Chandra} fields is imaged in \emph{VRI} and $H\alpha$ with
the Mosaic cameras on the KPNO- and CTIO-4m telescopes.  Ongoing efforts for complete
source classification include optical spectroscopy and infrared imaging of
heavily reddened Galactic Bulge regions.  A second major goal of ChaMPlane is to study
populations of stellar coronal sources.
By combining observations of coronal sources with spectral identifications from
ChaMPlane's optical survey, we will be able to constrain when stars develop coronae.

In this study, we determine the properties of the brightest serendipitous sources using
X-ray spectral fitting,
X-ray lightcurve analysis, and optical photometry of candidate counterparts (\S\ref{kp-bright-sample}).
We also test the predictions of quantile analysis for the first time (\S\ref{kp-qtile-analysis}).

\section{The bright sample\label{kp-bright-sample}}

\citet{kp_2005ApJ...635..920G} details the selection criteria for the entire ChaMPlane
Survey.  For this study, we apply further criteria,
including a requirement that the number of net source counts in the broad
(B$_{X}$, .3 -- 8 keV) band $\ga$ 250,
to confidently model and fit X-ray spectra.  We exclude the inner Bulge ($l > 358\deg$ and 
$l < 2\deg$) to study a less crowded environment.

Our bright sample is composed of 21 serendipitous X-ray sources.  Ten
have candidate optical counterparts with $R \la 24$; five of these have optical spectra in
our current spectral database.  Optical counterpart matching is in progress for three remaining
sources.

\subsection{The CV/qLMXB candidate\label{kp_analysis}}

\begin{figure}[!ht]
\begin{center}
\includegraphics{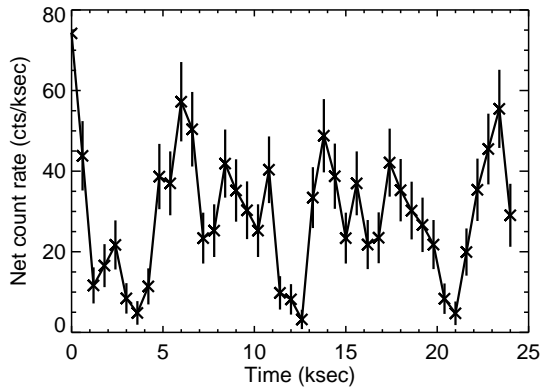}
\caption{Background subtracted X-ray lightcurve for the CV/qLMXB candidate, binned in
  600 second intervals.  A K-S test on the photon arrival times shows a significant
  (p = .009) departure from a constant count rate.\label{kp00090lc}}
\end{center}
\end{figure}

One X-ray source is well fit (reduced $\chi^{2} = .94$) by the absorbed powerlaw
model, with $\Gamma$ = 1.01$^{+.14}_{-.13}$ and
N$_{H}$ = .22$^{+.12}_{-.11} \times 10^{22}$ cm$^{-2}$.  The unabsorbed flux
in B$_{X}$ is $5.1 \pm 1.2 \times 10^{-13}$ erg cm$^{-2}$ s$^{-1}$.  Assuming the
Galactic dust distribution of \citet{kp_2001ApJ...556..181D}, the N$_{H}$
implies a distance of 1.3 kpc.  The unabsorbed luminosity in B$_{X}$ is
thus 10$^{32}$ erg s$^{-1}$, consistent with
luminosities of both qLMXBs and CVs.  The X-ray lightcurve 
(Fig. \ref{kp00090lc}) appears
periodic; it deviates from a constant count rate to 1\% significance.  Furthermore,
the candidate optical counterpart of this source
has a large H$\alpha$ excess ($H\alpha - R = -.48 \pm .13$) and is
bluer than many objects in the surrounding field.  The ratio of unabsorbed X-ray flux
to unabsorbed $R$ flux (i.e., unabsorbed F$_{X}$/F$_{R}$) is 28.3; for comparison, most stellar coronal
sources have unabsorbed F$_{X}$/F$_{R} \sim 10^{-3}$.  This source is most likely a 
CV or qLMXB.  Further work will help determine the precise nature of the compact object.

\subsection{The remaining sample}

\begin{figure}[!ht]
\begin{center}
\includegraphics{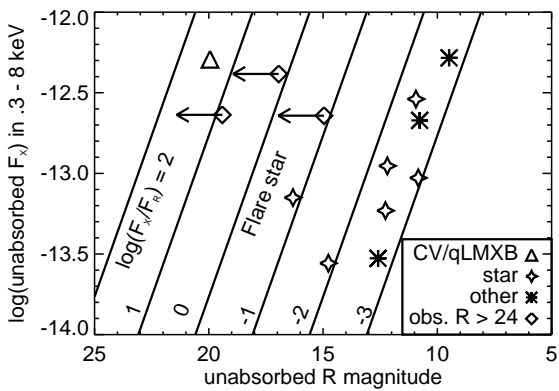}
\caption{Unabsorbed X-ray flux against unabsorbed $R$ magnitude for the sample.  Also shown
are lines of constant log(F$_{X}$/F$_{R}$).  Several X-ray sources have spectra fit well using
multiple models; only models with a strictly optimal reduced
$\chi^{2}$ are plotted.  Sources for which the optical counterpart is too faint are shown with
upper limits to the unabsorbed $R$ magnitude.\label{kp-sample-fxfr}}
\end{center}
\end{figure}

Combining X-ray spectral fits with candidate optical counterpart photometry and
spectroscopy in
the same way as in \S\ref{kp_analysis}, we characterize the remainder of our
sample.  Our preliminary results indicate the sample consists of
the CV/qLMXB candidate;  ten stellar sources, one of which shows an X-ray
flare during the \emph{Chandra} observation; three probable active
galactic nuclei (AGN); one candidate young stellar object (YSO);
and six unclassified sources.  These six include four sources with optical counterparts
too faint for ChaMPlane; if we assume a limiting absorbed
$R \simeq 24$, two of these four sources have unabsorbed F$_{X}$/F$_{R} > 1$,
values too high for the sources to be stellar coronae.  Unabsorbed F$_{X}$/F$_{R}$
for the sample is shown in Fig. \ref{kp-sample-fxfr}; the ratios range from $\sim 10^{-3}$ for
stellar coronal sources to $\simeq 30$ for the CV/qLMXB candidate.

\section{Quantile analysis\label{kp-qtile-analysis}}

\begin{figure}[!ht]
\begin{center}
\includegraphics{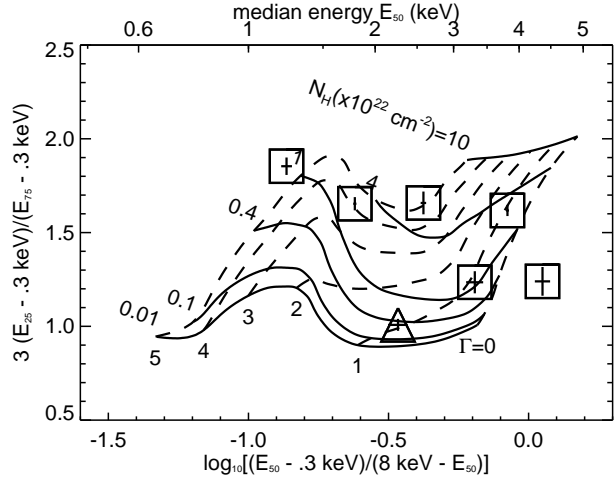}
\caption{Quantile color-color diagram (QCCD) for the absorbed powerlaw model.
Soft, unobscured sources are to the bottom-left, and hard, obscured sources are to the
top-right; spectral parameters (here, $\Gamma$) and N$_{H}$ are constrained by comparison
with model grids.
Plotted on this diagram are a few sources from our bright sample; the CV/qLMXB
candidate is marked with a triangle.  E$_{50}$ is the median energy;
E$_{25}$ and E$_{75}$ are the quartile energies.\label{kp_plaw_quantile}}
\end{center}
\end{figure}

Most X-ray sources in ChaMPlane's catalog do not
have enough counts for confident spectral fitting.  Quantile analysis, detailed in
\citet{kp_2004ApJ...614..508H}, is a method of constraining the absorption and
spectral parameters of low-count sources (see Fig. \ref{kp_plaw_quantile}).  We perform
quantile analysis for the 21 bright sources using both powerlaw and thermal
bremsstrahlung models.
For the first time, we test the predictions of quantile analysis against spectral
parameters from fitting.  We find that, of the 13 sources where the X-ray spectrum is
best fit by either a powerlaw or thermal bremsstrahlung model, only 2 have quantile values
outside the 1$\sigma$ results from spectral fitting.  All sources have quantile values
within the 2$\sigma$ results from spectral fitting.

\section{Conclusions}

Our population study of the brightest serendipitous X-ray sources in the ChaMPlane Survey
has uncovered a previously unreported CV/qLMXB,
ten stellar sources (including a flare star), three possible AGN, and
one candidate YSO.

We perform quantile analysis on our sample and find that it predicts spectral parameters and
absorption mostly to within the 1$\sigma$ errors of X-ray spectral fitting.

We plan to further study the CV/qLMXB candidate
and determine more properties (e.g. spectral types of stellar objects) of our bright X-ray
sample.

\acknowledgements We gratefully acknowledge the support of NSF grant 
AST-9731923.  This work was carried out as part of the Research Experience for
Undergraduates program at the Harvard-Smithsonian Center for Astrophysics.


\begin{thebibliography}{}
\bibitem[Warner(1995)]{kp_1995cvs..book.....W} Warner, B.\ 1995,
Cataclysmic Variable Stars (Cambridge: Cambridge University Press)
\bibitem[Grindlay et al.(2005)]{kp_2005ApJ...635..920G} Grindlay, J.~E., et 
al.\ 2005, \apj, 635, 920
\bibitem[Drimmel \& Spergel(2001)]{kp_2001ApJ...556..181D} Drimmel, R., \& 
Spergel, D.~N.\ 2001, \apj, 556, 181
\bibitem[Hong et al.(2004)]{kp_2004ApJ...614..508H} Hong, J., Schlegel, E.~M., 
\& Grindlay, J.~E.\ 2004, \apj, 614, 508
\end{thebibliography}
\end{document}